\documentclass[12pt]{article}
\usepackage{amssymb}
\begin{document}

\begin{flushright}
{\tt hep-th/0210101}
\end{flushright}

\vspace{5mm}

\begin{center}
{{\Large \bf Rolling Tachyons in String Cosmology}\\[12mm]
{Chanju Kim\footnote{cjkim@ewha.ac.kr}, 
Hang Bae Kim\footnote{HangBae.Kim@ipt.unil.ch},
Yoonbai Kim\footnote{yoonbai@skku.ac.kr}}\\[7mm]
{\it ${}^{1}$Department of Physics, Ewha Womans University}\\
{\it Seoul 120-750, Korea}\\[2mm]
{\it ${}^{2}$Institute of Theoretical Physics, University of Lausanne}\\
{\it CH-1015 Lausanne, Switzerland}\\[2mm]
{\it ${}^{3}$BK21 Physics Research Division and Institute of Basic Science\\
Sungkyunkwan University, Suwon 440-746, Korea}
}
\end{center}

\vspace{5mm}

\begin{abstract}
We study the role of rolling tachyons in the cosmological model with dilatonic
gravity. In the string frame, flat space solutions of both initial-stage and 
late-time are obtained in closed form.
In the Einstein frame, we show that every expanding solution is decelerating.
\end{abstract}


\newpage

\setcounter{equation}{0}
\section{\large\bf Introduction}

Though the situation seems not to be matured yet, {\it string cosmology} must 
be an intriguing subject to be tackled at every step of progress in both 
string theory and observational cosmology~\cite{SC}.
Recently, inspired by string theory D-branes and heterotic M theory,
brane cosmology attracted much attention.
Brane cosmology assumes that our universe starts out with branes embedded
in the higher dimensional spacetime, either stable or unstable.
Instability of D3-brane systems is represented by rolling of tachyon~\cite{Sen}
whose action is of Born-Infeld type~\cite{Gar} and has exponentially decaying 
potential~\cite{GS}. Since cosmological evolution of accelerating universe
was obtained at the early epoch of the rolling tachyon~\cite{Gib,Ale}, 
the cosmology involving the rolling tachyon attracts attention in relation with
various topics, e.g., inflation, dark matter, and reheating~\cite{tacos}, 
despite stringent difficulty in the simplest versions of this
theory~\cite{KL}. A nice aspect is 
the absence of open string excitations, which results in the absence of the
plane-wave solutions in effective field theory~\cite{Sen,IU} so that 
classical analysis of late-time based on homogeneity and isotropy may 
lead to a solid prediction.

When we deal with string cosmology, two indispensable fields are
gravitons and dilaton among various degrees of freedom.
For unstable branes, tachyonic degree of freedom should also be involved.
Thus, from the theory side of string cosmology with unstable branes,
the system of tachyon, dilaton, and gravitons provides a minimal setting.
In this paper, we consider the cosmology of unstable brane dominated by
tachyon and dilaton, based on the effective 4D action.
We assume that our brane is either a 3-brane whose transverse dimensions
are compactified or a space filling brane with compactified extra dimensions.
To be a realistic scenario leading to the final radiation, matter,
and dark energy dominated universes, other processes should also be clarified,
including the birth of (stable or unstable) branes,
the compactification of extra dimensions and the dilaton stabilization.
Though we need to understand physics 
behind them and their characteristic time scales to complete the brane 
world scenario, most of them rely on physics unknown at the moment.
Hence our purpose in this paper is restricted to the study of
intermediate stage dynamics when both the dilaton and the tachyon play
dominant roles in the evolution of the universe.
We do not introduce the stabilization potential for the dilaton,
and find the evolution of the scale factor and tachyon energy density
while both the dilaton and the tachyon are rolling.
Specifically, in the string frame, we obtain flat space solutions
at both the early epochs and late-times in closed form.
When transformed to the Einstein frame, analysis shows that every expanding
universe of the graviton-tachyon-dilaton system should be decelerating
irrespective of the specific shape of the tachyon potential.

The paper is organized as follows.
In section 2, we consider the effective action of a D3-brane system
involving graviton, dilaton, and tachyon in the string frame. 
For the late stage of tachyon-rolling, all possible cosmological solutions 
are obtained in closed form. 
In section 3, we reanalyze the system 
in the Einstein frame and show that every expanding universe is decelerating.
Section 4 is devoted to concluding remarks.

\section{Cosmological solutions in the string frame}

Let us begin with a cosmological model induced from string theory, which
is confined on a D3-brane of tension $T_{3}$ and
includes graviton $g_{\mu\nu}$, dilaton $\Phi$, and tachyon $T$.
In the string frame, the effective action of the bosonic sector of the D3-brane
system is given by \cite{Gar}
\begin{eqnarray}
S&=&\frac{1}{2\kappa^2}\int d^4x\sqrt{-g}\;e^{-2\Phi}\left(R
+4\nabla_\mu\Phi\nabla^\mu\Phi\right)
\nonumber\\
&&-T_{3}\int d^4x\;e^{-\Phi}\;V(T)
\sqrt{-\det (g_{\mu\nu}+\partial_{\mu}T\partial_{\nu}T)}\; ,
\label{act2}
\end{eqnarray}
where we turned off an Abelian gauge field on the D3-brane and
antisymmetric tensor fields of second rank both on the brane and
in the bulk, and set the vanishing cosmological constant.
Though there seems no consensus and no exact computation 
of tachyon potential $V(T)$ for all $T$'s, which measures change of the tension,
except for its maximum (max($V(T))=1$) and
minimum (min($V(T))=0$), its specific form can be chosen, for example, as an 
exponentially-decreasing potential 
\begin{equation}\label{V}
V(T)=\left\{
\begin{array}{ll}
\exp\left(-\frac{T^{2}}{8\ln 2}\right) & \mbox{for small $T$ but $T\ge 0$} \\
\exp\left(-\frac{T}{\sqrt{2}}\right) & \mbox{for large $T$}
\end{array}
,
\right.
\end{equation}
which connects two asymptotic expressions smoothly and is consistent with 
the results of superstring theory~\cite{Sen,GS}.

For cosmological solutions in the string frame, we try a spatially-homogeneous but 
time-dependent solution
\begin{equation}
ds^2=-dt^2+a^2(t)d\Omega^{2}_{k},\quad \Phi=\Phi(t),\quad T=T(t),
\label{cosmos1}
\end{equation}
where $d\Omega^{2}_{k}$ corresponds, at least locally, to the metric of 
$S^{3}$, $E^{3}$, or $H^{3}$ according to the value of $k=1,0,-1$, respectively.
{}From the action (\ref{act2}), 
we obtain the following equations
\begin{equation}
3\left(\frac{\dot a^2}{a^2}+\frac{k}{a^2}\right)
-2\left(3\frac{\dot a}{a}\dot\Phi-\dot\Phi^2\right) = 
\kappa^2 e^\Phi \rho_{T},
\label{metric3-1}
\end{equation}
\begin{equation}
2\frac{\ddot a}{a}+\frac{\dot a^2}{a^2}+\frac{k}{a^2}
-2\left(\ddot\Phi+2\frac{\dot a}{a}\dot\Phi-\dot\Phi^2\right) = 
-\kappa^2 e^\Phi p_{T},
\label{metric3-2}
\end{equation}
\begin{equation}
4\left(\ddot\Phi+3\frac{\dot a}{a}\dot\Phi-\dot\Phi^2\right)
-6\left(\frac{\ddot a}{a}+\frac{\dot a^2}{a^2}+\frac{k}{a^2}\right) = 
\kappa^2 e^\Phi p_{T},
\label{dilaton3}
\end{equation}
\begin{equation}
\frac{\ddot T}{1-\dot T^2}+\left(3\frac{\dot a}{a}-\dot\Phi\right)\dot T
+\frac{1}{V}\frac{dV}{dT}=0,
\label{tachyon3}
\end{equation}
where tachyon energy density $\rho_{T}$ and pressure $p_{T}$
defined by $T_{\;\;\;\;\nu}^{T\mu}\equiv
{\rm diag}(-\rho_{T},p_{T},p_{T},p_{T})$ are
\begin{equation}\label{pre}
\rho_T = T_{3}\frac{V(T)}{\sqrt{1-\dot T^2}}~~~{\rm and}
~~~p_T = -T_{3}V(T)\sqrt{1-\dot T^2}.
\end{equation}
The tachyon equation (\ref{tachyon3}) is equivalent to the following 
conservation equation
\begin{equation}
\dot\rho_T+(3H-\dot\Phi)\dot T^2\rho_T=0.
\label{conserv-eq}
\end{equation}
where $H=\dot a/a$ is the Hubble parameter.
In the absence of detailed knowledge of $V(T)$,
we will examine characters of the solutions
based on the simplicity of tachyon equation of state
\begin{equation}
p_T=w_T\rho_T,\quad w_T=\dot T^2-1.
\label{T-eq-of-state}
\end{equation}

By defining the shifted dilaton $\phi=2\Phi-3\ln a$,
we rewrite the equations (\ref{metric3-1})--(\ref{tachyon3}) as
\begin{eqnarray}
\dot\phi^2-3H^2 +6\frac{k}{a^{2}}
&=& 2\kappa^{2}e^{\frac{\phi}{2}}a^{\frac{3}{2}}\rho_T, \label{metric4-1} \\
2(\dot{H}-H\dot\phi) +4\frac{k}{a^{2}}
&=& \kappa^{2} e^{\frac{\phi}{2}}a^{\frac{3}{2}} p_T, \label{metric4-2} \\
\dot\phi^2-2\ddot\phi +3H^2 +6\frac{k}{a^{2}}
&=& -\kappa^{2} e^{\frac{\phi}{2}}a^{\frac{3}{2}} p_T, \label{dilaton4} \\
\frac{\ddot T}{1-\dot T^2}+\frac12 (3H-\dot\phi)\dot T
&=&-\frac{1}{V}\frac{dV}{dT} .
\label{tachyon4}
\end{eqnarray}
Note that $\sqrt{-g}$, or $a^3$ is not a scalar quantity
even in flat spatial geometry,
the shifted dilaton $\phi$ is not a scalar field in 3+1 dimensions.
The conservation equation (\ref{conserv-eq}) becomes
\begin{equation} \label{conserv-eq2}
\dot\rho_T + \frac12 (3H-\dot\phi)\dot T^2 \rho_T=0.
\end{equation}
Now Eqs.~(\ref{metric4-1})--(\ref{dilaton4}) and Eq.~(\ref{T-eq-of-state})
are summarized by the following two equations
\begin{eqnarray}
2\ddot\phi-\dot\phi^2+2H\dot\phi-3H^2-2\dot H &=& -10\frac{k}{a^{2}},
\label{xx1} \\
w_T\dot\phi^2+4H\dot\phi-3w_TH^2-4\dot H &=& -(8+6w_{T})\frac{k}{a^{2}}.
\label{xx2}
\end{eqnarray}
Let us consider only the flat metric $(k=0)$ in the rest part of the paper.
If we express the dilaton $\phi$ as a function of the scale factor $a(t)$,
$\phi(t)=\phi(a(t))$, we can introduce a new variable $\psi$ such as
\begin{equation}
\psi\equiv a\phi' =\frac{\dot{\phi}}{H},
\end{equation}
where the prime denotes the differentiation with respect to $a$, and the
second equality shows that $\psi$ is the ratio between $\dot{\phi}$ and $H$.
Then Eqs.~(\ref{xx1}) and (\ref{xx2}) are combined into a single first-order
differential equation for $\psi$ :
\begin{equation}\label{psi-eq}
4a\psi'+(\psi^2-3)(w_T\psi+2-w_T) = 0. 
\end{equation}

{}From now on we look for the solutions of Eq.~(\ref{psi-eq}).
Above all one may easily find a constant solution 
$\psi=\mp\sqrt{3}$ which is consistent with 
Eqs.~(\ref{metric4-1})--(\ref{dilaton4}) only when $\rho_T=0$~:
\begin{eqnarray}
a(t)&=&a_{0}(1\mp\sqrt{3}H_{0}t)^{\mp 1/\sqrt{3}},
~~~H(t)=\frac{H_{0}}{1\mp\sqrt{3}H_{0}t},\label{ata}\\
\Phi(t)&=&\Phi_{0}-\frac{1\pm\sqrt{3}}{2}\ln(1\mp\sqrt{3}H_{0}t),
\label{pta}
\end{eqnarray}
where $H_{0}=H(t=0)$, $a_{0}=a(t=0)$, and $\Phi_{0}=\Phi(t=0)$ 
throughout this section. However, exactly-vanishing tachyon density 
$\rho_{T}=0$ from Eq.~(\ref{metric4-1}) restricts 
strictly the validity range of this 
particular solution to that of vanishing tachyon potential, $V(T)=0$,
which leads to $T=\infty$.
The tachyon equation (\ref{tachyon4}) forces $\ddot{T}=0$ and $\dot{T}=1$ so
that the tachyon decouples $(w_{T}=p_{T}=0)$.  
Therefore, the obtained solution (\ref{ata})--(\ref{pta}) corresponds to 
that of string cosmology of the graviton and the dilaton before stabilization
but without the tachyon.

Since it is difficult to solve Eq.~(\ref{psi-eq}) with dynamical $w_{T}$,
let us assume that $w_T$ is time-independent
(or equivalently $a(t)$-independent).
We can think of the cases where
the constant $w_T$ can be a good approximation.
{}From the tachyon potential (\ref{V}),
the first case is onset of tachyon rolling around the maximum point
and the second case is late-time rolling at large $T$ region.
In fact we can demonstrate that those two cases are the only possibility
as far as no singularity evolves.

When $w_T$ is a nonzero constant, Eq.~(\ref{psi-eq}) allows
a particular solution
\begin{equation}
\psi=\frac{w_T-2}{w_T}\equiv\beta.
\end{equation}
This provides a consistent solution of Eqs.~(\ref{metric4-1})--(\ref{dilaton4})
\begin{eqnarray}
a(t) &=& a_0\left(1+\frac{w_T^2+2}{2w_T}H_0t\right)^{\frac{2w_T}{w_T^2+2}},
\qquad
H(t) =H_0\left(1+\frac{w_T^2+2}{2w_T}H_0t\right)^{-1}, 
\\
\Phi(t)&=&\Phi_{0}+\frac{2(2w_{T}-1)}{w^{2}_{T}+2}\ln
\left(1+\frac{w^{2}_{T}+2}{2w_{T}}H_{0}t\right). 
\label{ptw}
\end{eqnarray}
{}From Eq.~(\ref{metric4-1}) and Eq.~(\ref{conserv-eq2}), 
the tachyon energy density $\rho_{T}$ is given by
\begin{eqnarray}\label{rhos}
\rho_T(t) = \frac{2-2w_T-w_T^2}{w_T^2\kappa^2e^{\Phi_0}}H_{0}^{2}
\left(1+\frac{w_T^2+2}{2w_T}H_0 t\right)^{-\frac{2(1+w_T)^2}{w_T^2+2}} .
\end{eqnarray}
Since the obtained solution is a constant solution of $\psi$,
it has only three initially-undetermined constants.
Specifically, the solution should satisfy $\dot\Phi=[(2w_T-1)/w_T]H$
so that the initial conditions also satisfy a relation
$\dot\Phi_0=[(2w_T-1)/w_T]H_0$.
Once we assume general solutions of $a(t)$-dependent $\psi$
with keeping constant nonzero $w_{T}$, they should be classified
by four independent parameters $(a_{0},H_{0},\Phi_{0},\dot{\Phi}_{0})$
instead of three in Eq.~(\ref{ptw}).

According to the aforementioned condition for valid $w_{T}$ values,
the obtained solution in Eq.~(\ref{ptw}) may be physically
relevant as the onset solution of $w_{T}=-1$ $(\psi=3)$.
In this case, $\rho_T(t)$ is reduced to a constant 
$ \rho_T(t) = 3 e^{-\Phi_0} H_0^2 /\kappa^2 $.
Comparing this with the definition of $\rho_T$ in Eq.~(\ref{pre}),
the initial Hubble parameter $H_0$ is related to the dilaton as
$ H_0 = \pm \kappa e^{\Phi_0/2} \sqrt{T_3/3} $. Then, with $T(t) = 0$,
the tachyon equation of motion is automatically satisfied and hence
Eq.~(\ref{ptw}) becomes an exact solution of the
whole set of equations of motion (\ref{metric3-1})--(\ref{tachyon3}).
Since the tachyon field remains as constant at the maximum of the potential,
this solution describes the expanding or shrinking solution depending
on the initial Hubble parameter, with a constant vacuum energy corresponding
to brane tension due to tachyon sitting at the unstable 
equilibrium point.\footnote{The interpretation as expanding or
shrinking solution needs to be more careful,
since we are working in the string frame.
Actually the behaviors are reversed in the Einstein frame as we will
see in section 3.}

In order to study the behavior of the tachyon rolling down from the top of 
the potential, now we slightly perturb this solution, i.e., look for a
solution with nonzero but small $T$ dependence.
So we treat $T$ as a small expansion parameter
and work up to the first-order in $T$.
Since the unperturbed solution satisfies $ 3H = \dot\Phi $, 
the tachyon equation of motion (\ref{tachyon3}) becomes, to the 
first-order in $T$,
\begin{equation}
\ddot{T} = -\frac{1}{V} \frac{dV}{dT} .
\end{equation}
This can easily be integrated to
\begin{equation}\label{teqn}
\frac12\dot{T}^2 = -\ln V + \mbox{const} = \frac{T^2}{8\ln2} + \mbox{const},
\end{equation}
where we used the form of the potential near the origin (\ref{V}).
Given the initial condition that $T=T_0$ and $\dot T = 0$ at $t=0$,
we can solve this equation (\ref{teqn}) and obtain
\begin{equation} \label{perturbed}
T(t) = T_0 \cosh{\alpha t},
\end{equation}
where $\alpha = 1/2\sqrt{\ln2}$.
Therefore tachyon starts to roll down the potential as a hyperbolic
cosine function. 
Taking derivative, we find $\dot T = \alpha T_0 \sinh{\alpha t}$.
The range for which $\dot T$ remains small is then 
$t\lesssim t_r\equiv 2\sqrt{\ln2} \sinh^{-1}(2\sqrt{\ln2}/T_0)$,
during which the approximation $w_T \simeq -1$ is good.
Unless the initial value $T_0$ is fine-tuned,
the tachyon follows the onset solution (\ref{ptw})--(\ref{rhos})
for $t\lesssim t_r$ and enters into rolling mode.

For more general solutions, the first-order differential 
equation (\ref{psi-eq}) can be integrated to
\begin{equation}
a = C\left[\frac{\psi^2-3}{(\psi-\beta)^2}
\left(\frac{\psi-\sqrt3}{\psi+\sqrt3}\right)^{\frac{\beta}{\sqrt3}}
\right]^{\frac{\beta-1}{\beta^2-3}},
\end{equation}
where $C$ is an integration constant.
Note that this algebraic equation does not provide a closed form of $\psi$ 
in terms of the scale factor $a(t)$ except for a few cases, e.g., 
$w_T=0,-1/(\sqrt{3}-1/2),-2/(3\sqrt{3}-1)$.

Fortunately, for the late-time case of vanishing $w_T$,
we can obtain the solution in closed form
\begin{eqnarray}\label{cpm}
\psi = \sqrt3\ \frac
{C_+\left(\frac{a}{a_0}\right)^{\sqrt3/2}
+C_-\left(\frac{a}{a_0}\right)^{-\sqrt3/2}}
{C_+\left(\frac{a}{a_0}\right)^{\sqrt3/2}
-C_-\left(\frac{a}{a_0}\right)^{-\sqrt3/2}}.
\end{eqnarray}
Then the scale factor $a$ and the dilaton $\Phi$ are explicitly 
expressed as functions of time $t$ by solving the equations 
(\ref{metric4-2})--(\ref{dilaton4}):
\begin{eqnarray}
a(t) &=& a_0\left(\frac{C_-t+2}{C_+t+2}\right)^{1/\sqrt3},
\qquad H(t)=\frac{4H_0}{(C_-t+2)(C_+t+2)},
\label{att}\\
\Phi(t)&=&\Phi_0+\ln\left[
2\frac{(C_-t+2)^{(\sqrt3-1)/2}}{(C_+t+2)^{(\sqrt3+1)/2}}
\right],
\label{a-w0}
\end{eqnarray}
where $C_\pm=(3\mp\sqrt3)H_0-2\dot\Phi_0$.
We also read the tachyon density $\rho_T$ from Eq.~(\ref{metric4-1}) 
\begin{equation}\label{rhot}
\rho_T = C_+C_-e^{-\Phi_0}\ 
   \frac{(C_+t+2)^{(\sqrt3-1)/2}}{(C_-t+2)^{(\sqrt3+1)/2}}.
\end{equation}

Note that $C_{\pm}$ should have the same sign
from the positivity of the tachyon density (\ref{rhot}).
Let us first consider that both $C_{+}$ and $C_{-}$ are positive.
When $C_->C_+$ or equivalently $H_0>0$, the scale factor $a$ is growing
but saturates to a finite value
such as $a(\infty)=a_0 (C_-/C_+)^{1/\sqrt{3}}$ in the string frame.
When $C_-<C_+$, it decreases.
When $C_-=C_+$, $H_0=0$ so that the scale factor is a constant, $a(t)=a_0$.
For all of the cases, the dilaton $\Phi$ approaches negative infinity.
Note that $w_{T}=0$ means late-time, the tachyon density decreases like
$\rho_T\sim1/t$ as $t\rightarrow\infty$. Consistency check by using
Eq.~(\ref{conserv-eq2}) or equivalently by Eq.~(\ref{tachyon4}) provides us
the expected result, $\dot{T}\rightarrow1$.
If both $C_{+}$ and $C_{-}$ are negative, there appears a singularity at
finite time irrespective of relative magnitude of $C_{+}$ and $C_{-}$.

\section{Analysis in the Einstein frame}

In the previous section, it was possible to obtain the cosmological solutions
analytically in a few simple but physically meaningful limiting cases.
To study the physical implications of what we found,
however, we need to work in the Einstein frame.
In this section, we will convert the cosmological solutions obtained
in the string frame to those in the Einstein frame
and discuss the physical behaviors.
In the Einstein frame, the metric has the form
\footnote{In this section all the quantities are in the Einstein frame 
except the variables with subscript $s$ which denote the quantities 
in the string frame.}
\begin{equation} \label{einsteinmetric}
ds^2=e^{2\Phi}(-dt^2+a^2(t)d\Omega^{2}_{k}),
\end{equation}
and hence the time $t$ and the scale factor $a$ are related to
those in the string frame as
\begin{equation}\label{rel}
a_s=ae^{\Phi},\qquad dt_s=e^{\Phi}dt.
\end{equation}
Then the equations of motion for the flat
case $(k=0)$ in the string frame (\ref{metric3-1})--(\ref{tachyon3})
are converted to
\begin{eqnarray}
H^2 &=& \frac13 \dot\Phi^2 + \frac13 \kappa^2 e^{3\Phi}\rho_T\,, 
           \label{einstein5}\\
\dot H &=& - \dot\Phi^2 - \frac12 \kappa^2 e^{\Phi}\rho_T \dot T^2\,, 
           \label{einstein6}\\
\ddot \Phi + 3H\dot\Phi &=& - \frac12 \kappa^2 e^{3\Phi}(\rho_T - 2 p_T)\, ,
           \label{einstein7}
\end{eqnarray}
\begin{equation} \label{einstein4}
\frac{\ddot T}{1-e^{-2\Phi}\dot T^2}
+3H\dot T 
+\dot\Phi\dot T \frac{1-2e^{-2\Phi}\dot T^2}{1-e^{-2\Phi}\dot T^2}
+e^{2\Phi}\frac{1}{V}\frac{dV}{dT}=0.
\end{equation}
Tachyon energy density $\rho_T$ and pressure $p_T$ in the Einstein frame
are obtained by the replacement $\dot{T}_{s}=e^{-\Phi}\dot{T}$ 
in Eq.~(\ref{pre}) according to Eq.~(\ref{rel}),
and thereby $w_{T}$ is given as 
$w_{T}\equiv p_T /\rho_T=e^{-2\Phi}\dot T^2 -1$.
Demanding constant $w_{T}$ is nothing but asking a strong 
proportionality condition between the dilaton and tachyon, 
$\dot{T}\propto e^{\Phi}$.
Note that the pressure $p_{T}$ as shown in Eq.~(\ref{pre}) is always 
negative irrespective 
of both specific form of the tachyon potential $(V(T)\ge 0)$ and the value of
the kinetic term $(e^{-2\Phi}\dot{T}^{2}\le 1)$, and the value of $w_T$
interpolates smoothly between $-1$ and 0.

First we observe that the right-hand side of Eq.~(\ref{einstein5}) is always 
positive, which means that the Hubble parameter $H(t)$ is either 
positive definite or negative definite for all $t$ and it cannot change 
the sign in the Einstein frame. 
Obviously it is a natural consequence of the weak energy condition.
Let us first consider the case of positive Hubble parameter, $H(t)>0$.  
Eq.~(\ref{einstein6}) shows $\dot H$ consists of two
terms both of which are negative definite for all $t$.  Since $H>0$ by
assumption, the only consistent behavior of $H$ in this case is that
$\dot H$ vanishes as $t\rightarrow \infty$, which, in turn implies that
$\dot\Phi$ and $e^{\Phi}\rho_T \dot T^2$ go to zero, separately.
It also means that $H$ should be a regular function for all $t$. In order to
find the large $t$ behavior of $H(t)$, one has to study $e^{-\Phi} \dot T$
in large $t$ limit which appears in the definition of $w_T$ in the
Einstein frame. Knowing that the functions are regular, it is not difficult
to show that the only possible behavior is $e^{-\Phi} \dot T \rightarrow 1$
as $t \rightarrow \infty$ after some straightforward analysis of
Eqs.~(\ref{einstein5})--(\ref{einstein7}). Combining it with the fact that 
$\dot\Phi$ and $e^{\Phi}\rho_T \dot T^2$ vanish, we can immediately 
conclude from Eq.~(\ref{einstein5}) that $H(t)$ should go to zero in large 
$t$ limit.

The asymptotic behavior of fields in case of the positive Hubble parameter
can be found from the solution (\ref{a-w0}) since $w_T$ is essentially
zero for large $t$ as we just have seen above. The only thing to do is 
to transform the expressions in the string frame to those in the 
Einstein frame, using the relation (\ref{rel}).
Therefore, for large $t$, we find
\begin{eqnarray} \label{einstein_a}
a(t_s) &=& a_s(t_s)e^{-\Phi(t_s)}
\simeq\frac{1}{2}a_{s0}e^{-\Phi_{0}}
(C_{+}t_s+2)^{\frac{\sqrt3+1}{2\sqrt3}}
(C_{-}t_s+2)^{\frac{\sqrt3-1}{2\sqrt3}}, \nonumber \\
t &=& \int dt_s e^{-\Phi} 
    \simeq 2 e^{-\Phi_0} \int dt_s 
       \frac{(C_+ t_s+2)^{(\sqrt3+1)/2}}{(C_- t_s+2)^{(\sqrt3-1)/2}}.
\end{eqnarray}
One can also identify the initial Hubble parameter $H_0$ in terms of $C_\pm$ as
\begin{equation}
H_0 = \frac14 e^{\Phi_0}\left[
               \left( 1 - \frac1{\sqrt3} \right) C_{-} 
             + \left( 1 + \frac1{\sqrt3} \right) C_{+} \right].
\end{equation}
Note that $C_\pm$ have the same sign as the Hubble parameter $H$.
Now, with $C_\pm>0$, one can easily confirm from the above equation 
(\ref{einstein_a}) that all the functions indeed behave regularly.
In $t_s\rightarrow\infty$ limit, $a\sim t_s$ and $t \sim t_s^{2}$
so that the asymptotic behavior of the scale factor becomes $a\sim t^{1/2}$.
This power law expansion in flat space is contrasted with the result of
Einstein gravity without the dilaton $\Phi$, where ultimately the scale
factor ceases to increase, $\lim_{t\rightarrow\infty}a(t)\rightarrow$ constant.
The behavior of tachyon density $\rho_T$ can be read from
Eq.~(\ref{rhot}) with $t$ replaced by $t_s$, which shows that 
$\rho_T \sim t^{-1/2}$.
Since $w_T$ also goes to zero, the fluid of condensed tachyon becomes 
pressureless.
Differently from ordinary scalar matter where matter domination 
of pressureless gas is achieved
for the minimum kinetic energy $(\dot{T}\rightarrow 0)$, it is
attained for the maximum value of time dependence $(e^{-\Phi}\dot{T}
\rightarrow 1$ as $T\rightarrow \infty)$ for the tachyon potential 
given in Eq.~(\ref{V}).

When the Hubble parameter $H$ is negative, the situation is a bit more 
complicated. Since $\dot H<0$ always,
$H$ becomes more and more negative and there is a possibility that
eventually $H$ diverges to negative infinity at some finite time.
Indeed, it turns out that all solutions in this case develop a singularity
at some finite time at which $H \rightarrow -\infty$ and $a \rightarrow 0$.
These big crunch solutions may not describe viable universes
in the sense of observed cosmological data.
Depending on initial conditions, the dilaton $\Phi$ diverges to either $\infty$ or
$-\infty$ and $\dot T$ goes to either $\infty$ or zero with the factor
$e^{-\Phi} \dot T$ remaining finite. It is rather tedious and not much
illuminating to show this explicitly, so here we will just content
ourselves to present a simple argument to understand the behavior. 
Since the tachyon field $T$ rolls down from the maximum of the potential 
to the minimum at infinite $T$, it is physically clear 
that $ \dot T_s = e^{-\Phi} \dot T$ would eventually go to one unless
there is a singularity at some finite time. Suppose that there appeared
no singularity until some long time had passed so that $e^{-\Phi} \dot T$ 
approached to one sufficiently closely. Then Eq.~(\ref{einstein_a}) should be 
a good approximate solution in this case. However, we know that both 
$C_\pm$ are negative when $H<0$ and Eq.~(\ref{einstein_a}) is clearly
singular in this case. We have also verified the singular behavior for 
various initial conditions using numerical analysis.

As mentioned in the previous section, the tachyon $T$ is decoupled when 
$e^{-\Phi}\dot{T}=1$ and $T=\infty$. In this decoupling limit, characters of
the Einstein equations (\ref{einstein5})--(\ref{einstein6}) that $H^{2}>0$
and $\dot{H}<0$ do not change so that all the previous arguments can be applied.
Well-known cosmological solution of the dilaton gravity before stabilization
of the dilaton is
\begin{eqnarray}
a(t)&=&a_0 (1+3H_{0}t)^{1/3},~~~H(t)=\frac{H_{0}}{1+3H_{0}t},
\label{adg}\\
\Phi(t)&=&\Phi_{0}\pm\frac{1}{\sqrt{3}}\ln(1+3H_{0}t),
\label{pdg}
\end{eqnarray}
where the $(\pm)$ sign in Eq.~(\ref{pdg}) is due to the reflection symmetry
$(\Phi\leftrightarrow -\Phi)$ in the equations  
(\ref{einstein5})--(\ref{einstein7}).
This solution can also be obtained throughout a transformation (\ref{rel})
from Eqs.~(\ref{ata})--(\ref{pta}).
For $H_{0}<0$, it is a big crunch solution $(a\rightarrow 0)$ which encounters
singularity $(H\rightarrow\infty,~\Phi\rightarrow \mp\infty)$ as $t\rightarrow
1/3|H_0|$. For $H_{0}>0$, it is an expanding but decelerating solution.
Since $a\sim t^{1/3}$, the power of expansion rate is increased from $1/3$ to
$1/2$ by the tachyonic effect as expected.

So far we discussed generic properties and asymptotic behaviors of solutions 
in the Einstein frame. Now we consider the behavior at the onset.
The solution (\ref{ptw}) obtained by assuming constant $w_T$ is
transformed to the Einstein frame as
\begin{eqnarray}
a(t) &=& a_0\left[1+\frac{(w_T-2)^2}{2(1-w_T)}H_0t
        \right]^{\frac{2(1-w_T)}{(w_T-2)^2}}, \nonumber \\
e^{\Phi(t)} &=& e^{\Phi_0}\left[1+\frac{(w_T-2)^2}{2(1-w_T)}H_0t
        \right]^{\frac{2(2w_T-1)}{(w_T-2)^2}},
\end{eqnarray}
where the initial Hubble parameter $H_0$ is related to that in the string
frame by $H_0=e^{\Phi_0}H_{0s}(1-w_T)/w_T$.
Note that $H_0$ and $H_{0s}$ have opposite signs since $w_T<0$. Therefore
the expanding (shrinking) solution in the string frame corresponds to the 
shrinking (expanding) solution in the Einstein frame.
For the onset solution with $w_T=-1$, the tachyon energy density $\rho_T$
is a constant as before, $\rho_T(t) = 3e^{-3\Phi_0} H_0^2/4 \kappa^2$. 
Then the initial Hubble parameter is given by 
$H_0 = \pm 2\kappa e^{3\Phi_0/2} \sqrt{T_3/3}$, which describes the exact
solution that tachyon remains at the origin as explained in section 2.
Under a small perturbation tachyon starts
rolling down according to Eq.~(\ref{perturbed}) with $t$ replaced by
$t_s$. The rest of the discussion on the rolling behavior is the same as 
in the string frame and the details will not be repeated here.

In conclusion the cosmological solution can be classified into two categories
depending on the value of the Hubble parameter $H(t)$ in the Einstein frame.
When the initial Hubble parameter $H_0$ is positive, the solution is 
regular and the universe is expanding but decelerating as
$a(t) \sim \sqrt{t}$ while $e^{\Phi(t)}$ vanishes. When $H_0$ is negative, 
there appears a singularity at some finite time $t$ at which the universe 
shrinks to zero.

\section{Concluding remarks}

In this paper we have discussed cosmological solutions of the 
effective theory of rolling tachyon coupled to gravitons and dilaton.
In the study of homogeneous and isotropic universes, we found 
initial-stage and late-time solutions in closed form in addition to
the known solution of the gravitons and dilaton in the decoupling 
limit of the tachyon. Those obtained solutions included 
expanding universes, big crunch solutions,
and even the static universe in the string frame. 
The dilatonic part of those also indicated that
the dilaton could not be stabilized in the system of our interest
without the dilaton potential.

We have also provided a description of cosmological solutions in 
the Einstein frame. Einstein equations were summarized by the 
positivity of square of the Hubble parameter and the negativity 
of derivative of the Hubble parameter irrespective of
specific shape of the rolling tachyon potential. This implies
that once the universe starts expanding, then it continues
expanding eternally but decelerating.

We conclude with the list of intriguing questions for further study.
One would like to generalize this to that with 
the Abelian gauge field on the brane~\cite{MW} and ask whether or not
radiation dominated era is possible.
As was the original string cosmology, stabilization mechanism of dilaton
should be understood and its positive effect to our unsatisfactory 
decelerating universe solutions is carefully investigated~\cite{DGM}.
Various topics for the rolling tachyons asked in the Einstein gravity 
should be addressed again in the context of dilaton gravity 
such as existence of inflationary era, possibility as a source of 
quintessence, reheating without oscillating tachyon modes, 
cosmological perturbation and structure formation, and so on.

\section*{Acknowledgements}
We would like to thank Nakwoo Kim and Jung Tay Yee for discussions on D-brane
systems. H.~B.~Kim and Y.~Kim would like to acknowledge the hospitality 
of Korea Institute for Advanced Study where a part of this work has been done.
This work was supported by Korea Research Foundation Grant 
KRF-2002-070-C00025(C.K.) and the Swiss Science Foundation,
grant 21-58947.99(H.B.K.), and is
the result of research activities (Astrophysical Research
Center for the Structure and Evolution of the Cosmos (ARCSEC) and
the Basic Research Program, R01-2000-000-00021-0)
supported by Korea Science $\&$ Engineering Foundation(Y.K.).

\end{document}